\documentclass[a4paper,12pt]{article}
\usepackage[]{amsmath,amssymb}
\usepackage{amsfonts}
\usepackage[dvips]{graphicx}
\usepackage{upgreek}
\usepackage{array}
\usepackage{color}
\title{All-loop calculation of the Reggeon field theory amplitudes via stochastic 
model}

\author{R.S. Kolevatov$^{1,3}$\thanks{{rodion.kolevatov@fys.uio.no}}\quad
K.G. Boreskov$^{2}$\thanks{{boreskov@itep.ru}}\quad
L.V. Bravina$^{1}$\thanks{{larissa.bravina@fys.uio.no}} \and
\raggedright
\small $^1$Department of Physics, University of Oslo, PB1048 Blindern, N-0316 Oslo, Norway \and
\small $^2$Institute of Theoretical and Experimental Physics, 117259, Moscow, Russia\phantom{000000.}\and
\small $^3$Department of High Energy Physics, Saint-Petersburg State University,\phantom{0000000000}\\
\small Ulyanovskaya 1, 198504 Saint-Petersburg, Russia\phantom{0000000000000000000000000000.}
}

\date{}
\def\b{{\bf b}}

\def\z{{\bf z}}
\def\q{{\bf q}}

\def\B{{\cal B}}

\def\Z{{\cal Z}}
\def\X{{\cal X}}
\def\Y{{\cal Y}}

\def\p+{\! + \!}
\def\m-{\! - \!}
\def\r={\! = \!}
\def\Im{{\mathrm{Im}\,}}

\def\scs{\scriptstyle}
\def\scss{\scriptscriptstyle}
\def\N{{\scss N}}

\def\phid{\phi^\dagger}
\def\r3P{r_{{\scss 3P}}}
\def\Aph{A\phantom{\widetilde{A}\!\!\!\!}}
\def\GeV{~\mathrm{GeV}}
\def\fm{~\mathrm{fm}}

\begin{document}

\maketitle

\abstract{
The evolution equations for Green functions of the Reggeon Field Theory (RFT) are equivalent to
those of the inclusive distributions for the reaction-diffusion system of classical particles. We
use this equivalence to obtain numerically Green functions and amplitudes of the RFT with all loop
contributions included. The numerical realization of the approach is described and some important
applications including total and elastic proton--proton cross sections are studied. 
It is shown that the loop diagram contribution is essential but can be imitated in the eikonal cross section
description by changing the Pomeron intercept. A role of the quartic Pomeron coupling which is an
inherent part of the stochastic model is shown to be negligible for available energies.
%\PACS{{12.40.Nn}{Regge theory, duality, absorptive/optical models} \and
%     {13.85.-t}{Hadron-induced high- and super-high-energy interactions (energy $>$ 10 GeV)} \and
%     {13.85.Lg}{Total cross sections}}
%\keywords{Reggeon Field theory, Pomeron loops, reaction-diffusion models, total cross sections, elastic cross
%sections.}
}

\normalsize

\section{Introduction.}

The Reggeon Field Theory (RFT) \cite{Gribov'64}  is successfully used up to nowadays for describing
the soft part of the strong interaction dynamics at high energies. 
It is based on the very general principles which should obey $s$-channel amplitudes such as  analiticity, crossing and $t$-channel unitarity. Particular field theories like QCD should obey the same principles. The Regge-like behavior was observed in QCD and some of the RFT components which are not deducible from the first principles like the Pomeron intercept were obtained within certain approximations. 
These approximations of the QCD however are questionable to apply to the processes which do not involve high momentum transfer. Moreover, it is rather complicated to fulfil the reqirements of full many-particle  $t$-channel unitarity which is very important for high-energy dynamics. Due to the generality and phenomenological success of the RFT in many aspects it can be viewed as complementary to the QCD and should be treated on the equal grounds with the QCD-induced models.

For the description of the experimental data in the reggeon approach one usually uses the simplest reggeon diagrams corresponding to the pomeron pole and a set of nonenhanced regge-cut graphs providing $s$-channel unitarity. However for the case of the supercritical pomeron (with the intercept larger than one) the interaction between pomerons may be essential and can become especially important at the LHC energies so the contribution of loop reggeon diagrams should be estimated. 
In this paper we present such estimates for the RFT with triple and quartic pomeron vertices using the mathematical equivalence  of this theory to dynamics of stochastic system of classical particles \cite{Grassberger:1978pr}.

The Lagrangian density of the
Reggeon Field Theory is formulated by using the non-relativistic Pomeron field $\phi(y,\b)$
\begin{equation}
 \mathcal{L} = \frac{1}{2} \phi^\dagger (\overleftarrow{\partial_y} - \overrightarrow{\partial_y})\phi
 - \alpha' (\nabla_\b \phid)(\nabla_\b \phi)   + \Delta \phid\phi + \mathcal{L}_{int} ~,
\label{eq:LRFT}
\end{equation}
where the rapidity $y$ plays a role of time, $\b$ represents position in the impact parameter
plane, parameters $\alpha'$ and $\Delta$ describe the linear Pomeron trajectory,
$\alpha_P(t)=(1+\Delta)+\alpha' t$, and $L_{int}$ contains the interaction terms. The usual choice
for the interaction terms is the option when it generates only triple Pomeron vertices, however
terms of higher order in $\phi$ and $\phi^\dagger$ can also enter.

It was shown in \cite{Grassberger:1978pr} and discussed in the subsequent papers  \cite{Boreskov:2001nw,Bondarenko'06} that the
classical stochastic system of particles (called ``partons'' for brevity\footnote{{We note that these classical partons should not be confused with the partons in the Fock wave function or the color dipoles in the QCD induced cascade models \cite{Salam:1995uy,Avsar:2006jy}.}}) which undergo diffusion
in 2-dimensional space and interact in a specific way (the particles can split, die, fuse or
annihilate in pairs) allows a field-theoretical realization corresponding to the RFT Lagrangian
\eqref{eq:LRFT} with
\begin{equation}
 \mathcal{L}_{int} = i\,\r3P \phi^\dagger \phi (\phi^\dagger + \phi) + \chi {\phi^\dagger}^2 \phi^2 +
 j_0 \phi^\dagger + j_Y \phi ~.
\label{L_int}
\end{equation}
The interaction includes both triple (nonhermitian) and quartic terms, the source term $j_0(0,\b)$
gives the initial condition for the $y$-evolution (``parton'' distribution in the projectile) and
the term $j_Y(Y,\b)$ describes the partonic structure of the target and is necessary for
calculating the scattering amplitude at total rapidity $Y$.
The values of the Reggeon couplings in \eqref{L_int} are connected with the rate parameters of the
stochastic model. Equality of couplings $\r3P$ for both triple terms in \eqref{L_int} imposed by
Lorentz invariance requires tuning of the parameters.
The multiparticle inclusive distributions of the stochastic system turn out to be
identical to the Green functions of the RFT. This correspondence may be used to compute
the amplitudes of that specific model of the Reggeon field theory with the account of all
loops.

%\textcolor{magenta}{The reaction-diffusion approach to the RFT 
%has a formal resembalnce to a certain numerical approaches to 
%calculation of the amplitude in terms of the dipole model [Mueller].
%There the amplitude is produced via linkage of dipole cascades 
%which develop independently from projectile and target [Salam???, Lund]. 
%In spite of the formal similiarity it is not at all straightforward, 
%if possible, to our mind to establish the correspondence between the two 
%calculation schemes. At the same time, duality between the stochastic 
%reaction-diffusion system and the RFT is exact which allows the numerical 
%calculations of the RFT amplitudes which take into account interactions 
%into all orders of the RFT coupling constant and number of loops.}

In this paper we present a numerical realization of the stochastic approach and give some estimates
of the loop diagram contribution for the RFT in $d=0$ and $d=2$ dimensions.
 In the next section we recall the notations and the definitions of the stochastic
approach and outline the correspondence between the parameters of the RFT and that of the
stochastic approach. Section 3 is devoted to the description of the numerical method used for the
computations of the amplitude. In Section 4 we analyze asymptotic behaviour of the Pomeron Green
function in $d=0$ and $d=2$ dimensions for various sets of parameters. The role of the quartic
Pomeron coupling is discussed. In Section 5 the correspondence between the parameters of the stochastic
model and the couplings of the Reggeon field theory is established by comparing contributions of
simplest diagrams. %The restriction to the model algorithm are discussed which allow to
%calculate contribution of limited set of reggeon diagrams in the (quasi)eikonal  approximation.
We also discuss the restrictions which one must impose on the numerical algorithm to reproduce the 
contribution of the limited sets of the Reggeon diagrams which enter in the (quasi)eikonal approximation.
Then the calculated contribution to proton-proton cross sections of the full set of the Reggeon diagrams is
compared to the standard calculations in the quasieikonal approximation and the role of the Pomeron
interactions is estimated. The effective change of the Pomeron intercept due to account of these
interactions is found. Concluding remarks are presented in the last section.

\section{The stochastic model}
\label{sec2}

Here we briefly describe the stochastic model of the RFT (also known as reaction-diffusion
approach) as it is formulated in \cite{Boreskov:2001nw}. We follow the evolution of the system of
classic particles, ``partons'', which interact in a certain way. The partons are allowed to move
randomly in the transverse plane, and this diffusive motion is described by a diffusion coefficient
$D$. The particles can split, $A\rightarrow A+A$, with a splitting probability per unit time
$\lambda$, or die, $A\rightarrow \emptyset$, with a death probability $m_1$. When two partons are
brought within the reaction range due to the diffusion, they can pairwise fuse, $A+A\rightarrow A$, or
annihilate, $A+A\rightarrow \emptyset$. To describe the rate of these processes we use dimensional
fusion and annihilation constants, $\sigma_\nu$ and $\sigma_{m_2}$:
\begin{align}
 \sigma_\nu \equiv \int d^2b\, p_{\nu}(b) ~, \quad
 \sigma_{m_2} \equiv \int d^2b\, p_{m_2}(b) ~.
\end{align}
The dimension of $\sigma_{m_2}$ and $\sigma_{\nu}$ thus is length squared. The characteristic
scale, that is the effective interaction distance, is determined by the functions $p_\nu(b)$ and
$p_{m_2}(b)$. It is assumed to be small compared to the characteristic scale of the problem (e.g.
proton radius).%
\footnote{~Note that in ref.\cite{Boreskov:2001nw} the fusion and annihilation constants (denoted
there as $\nu$ and $m_2$ respectively) are defined with respect to a single parton and thus are
twice less compared to $\sigma_\nu$ and $\sigma_{m_2}$.
}

Throughout the evolution the system of partons is described by a set of symmetrized probability
densities $\rho_\N(y;\b_1,\ldots, \b_N)$ of finding exactly $N$ partons at evolution time $y$ at
positions $\{\b_1,\ldots, \b_N\}$ in the transverse plane. For brevity we use the notation $\B_N$
for the set $\{\b_1,\ldots, \b_N\}$ and $d\B_N$ for
 $d{\bf b}_1 \ldots d{\bf b}_N$. The probability densities are normalized according to
\begin{equation}
\label{norm_rho}
\frac{1}{N!}\int d\B_\N \rho_\N (y;\B_\N) = p_\N (y) ,
\end{equation}
where $p_N(y)$ is the probability of having exactly $N$ partons at the evolution time $y$,
$\sum p_N(y) =1$.

The important quantities which have a direct analog in the Reggeon field theory are the inclusive
$s$-parton distributions, which correspond to observing $s$ partons at positions
$\Z_s=\{\z_1,\ldots, \z_s\}$ and integrating out all the rest. They are defined as
\begin{equation}
 f_s(y;\Z_s)= \sum_N \cfrac{1}{(N-s)!} \int\! d\B_N \;\rho_\N(y;\B_N)
 \prod_{i=1}^{s} \delta(\z_i-\b_i). \label{def_f}
\end{equation}
Due to the symmetry properties of $\rho_N(y; \B_N)$ the inclusive distributions $f_s(y;\Z_s)$ are
also symmetric functions of their variables. As follows from the normalization of the $\rho_\N$,
the inclusive distributions are normalized to the factorial moments $\mu_s$ of the distribution
$p_\N(y)$:
\begin{equation}
\int d\Z_s f_s(y;\Z_s)  = \sum \frac{N!}{(N-s)!}\, p_N (y) \equiv \mu_s(y).
\end{equation}

As it is pointed out in \cite{Boreskov:2001nw} the evolution equations in time $y$ for the inclusive
distributions~$f_s(y; \Z_s)$ 
\begin{multline}
\frac{d}{dy}  f_s (y; \Z_s) = \, D\, {\nabla}_s^2 f_s (y; \Z_s) \p +  (\lambda - m_1) s  f_s (y; \Z_s) +
\\ \phantom{\frac{d}{dy}}
\p+ \lambda\! \sum_{k,l=1}^{s\geqslant  2} f_{s-1} (y;\Z_s^{(l)}) \delta (\z_k -\z_l)
\m- \nu\! \sum_{k,l=1}^{s\geqslant  2} f_s (y; \Z_s) \delta (\z_k - \z_l) -
\\ \phantom{\frac{d}{dy}}
\m- (2 m_2 +\nu )\!  \sum_{k=1}^s f_{s+1} (y; \Z_s^{(k)},\z_k,\z_k) ~, \qquad (s=1,2,\ldots) 
\end{multline}
\noindent exactly reproduce the evolution equations in rapidity for the Green functions of the RFT with
the interaction term \eqref{L_int}.
This means that the inclusive distributions themselves are proportional to the Green functions of
the RFT %averaged over the number and initial positions of pomerons at $y=0$:
in convolution with the particle--$m$-Pomeron vertices of the RFT $\mathcal{N}_m(\X_m)$: 
\begin{equation}
 f_s(y;\Z_s)  \propto
%=  \epsilon^{s/2} 
\sum_m \int d\X_m\, \mathcal{N}_m(\X_m) G_{mn}(0;\X_m|y;\Z_n) ~. \label{fs-vertex}
\end{equation}
where the Green functions $G_{mn}(0;\X_m|y;\Z_n)$ describe the transition of $m$ Pomerons at rapidity $y=0$ into $n$ Pomerons at rapidity $y$.
The proportionality factor takes into account the dimension of the inclusive distributions and
should be specified at calculating Regge amplitudes (see below).
 The vertices $\mathcal{N}_m(\X_m)$ determine the initial distribution of partons in number and
transverse positions at zero time $y=0$ which are used as a starting point for the evolution of
$f_s$. We will attend to this point in section \ref{sec5}.

The amplitude of the RFT as a function of the total rapidity $Y$ and impact
parameter $\b$ is expressed in terms of the set of the inclusive partonic distributions
$f_s(y,\Z_s)$ for the projectile $A$ and $\tilde{f_s}(Y-y,\tilde \Z_s)$ for the target $\tilde{A}$
taken at arbitrary intermediate rapidity $y$ through the procedure of linkage:
\begin{multline}
 T(Y) = \langle A | T | \tilde{A} \rangle 
 =\sum_{s=1}^\infty \frac{(-1)^{s-1}} {s!} \int d\Z_s
 d\tilde \Z_s f_s(y;\Z_s ) \tilde f_s(Y-y;\tilde \Z_s) \prod_{i=1}^s g ({\bf z}_i - \tilde {\bf z}_i - {\bf b}) .
\label{TST}
\end{multline}
This procedure realizes a linkage of partons (Pomerons) of the target and the projectile through a
convolution with the linking functions $g(\ldots)$. These functions are narrow functions of the
$\delta$-function type%
 \footnote{~ The original paper \cite{Boreskov:2001nw} had
 $\epsilon^s \prod \delta^{(2)} (z_i - \tilde z_i - \b)$  in (\ref{TST}) instead of more general expression
$\prod g (z_i - \tilde z_i - {\bf b})$, however this restriction is not essential and has mainly
the computational significance. The only important thing is that the characteristic scale
$\epsilon$ of the functions $g$ must be small compared to the scale of the problem. It differs, in
general, from scales of the functions $p_{\nu}$ and $p_{m_2}$.}
 normalized to some parton size $\epsilon =\pi a^2$. They will be
specified in section \ref{sec3}.

The formula \eqref{TST} links two partonic cascades coming from the target and the projectile at
the rapidity value $y$.
 It was proved in \cite{Boreskov:2001nw} that, provided the scales of the functions
$g(b)$, $p_{\nu}(b)$ and $p_{m_2}(b)$ are small, the convolution (\ref{TST}) depends only on the
overall rapidity/evolution time $Y$ but not on the position of the linking point $y$ if a certain condition on the
model parameters is satisfied. Namely,
\begin{equation}
 \lambda \int g(b) db = \int p_{m_2}(b) db +\frac{1}{2}\int p_{\nu}{b} db ~,
\label{inv_cond}
\end{equation}
or,
\begin{align*}
 \lambda \epsilon = \sigma_{m_2} +\frac{1}{2} \sigma_{\nu} ~.
\end{align*}
 This condition in fact corresponds to the equality of the Pomeron fusion and
splitting vertex.

The relation between the Reggeon field theory parameters and the ones of the stochastic model is
given in table \ref{RFT-ST-rel}.
 \extrarowheight=3pt
\begin{table}[h]
 \centering
 \caption{Relation between the parameters of the RFT and those of the stochastic approach.}
 \vskip 2mm
\begin{tabular}{|c|c|}
\hline
 \multicolumn{1}{|c|}{RFT} & stochastic model\\
 \hline
Rapidity $y$ & Evolution time $y$ \\
Slope $\alpha'$ & Diffusion coefficient $D$\\
$\Delta = \alpha(0)-1$ & $\lambda - m_1$ \\
Splitting vertex $\r3P$ & $\lambda \sqrt{\epsilon}$\\
Fusion vertex $\r3P$ & $(\sigma_{m_2}+\tfrac{1}{2}\sigma_{\nu})/\sqrt{\epsilon}$ \\
Quartic coupling $\chi$ & $\tfrac{1}{2} (\sigma_{m_2}+\sigma_{\nu}) $ \\[2mm]
 \hline
\end{tabular}
\label{RFT-ST-rel}
\end{table}

Since the inclusive distributions (\ref{def_f}) are in direct correspondence with the exact Green
functions of the RFT with the account of all loops (the model characteristic scale of $p_{\nu}$ and
$p_{m_2}$ plays a role of the cutoff for the RFT Green functions), the equation (\ref{TST}) can be
used for obtaining the exact numerical answer for the RFT scattering amplitude.

\section{Numerical method.}
\label{sec3}

The inclusive distributions $f_s$ and $\tilde f_s$ which enter (\ref{TST}) are defined by the
evolution of the partonic subsystems with a certain initial distribution of partons in their number and
transverse positions. Basically one may assume a straightforward way of calculating the interaction
amplitude according to (\ref{TST}) by making a Monte-Carlo average over the initial conditions
and computing $f_s$ and $\tilde f_s$ explicitly. This method, however, seems to be numerically
unaffordable. For example, besides the evolution of partonic system and computing the distributions
$f_s$ on some grid which one has to introduce, it is necessary to perform symmetrization of the
$f_s$'s which (provided the grid has fairly small cells) is extremely time consuming, or,
alternatively, increase the number of Monte-Carlo simulations in the same proportion. The
corresponding number of operations can be estimated as $M!/(M-s)!$, which is the number of choices
of $s$ out of $M$ points of the grid together with number of transpositions.

Let's rewrite (\ref{TST}) explicitly indicating the averaging over the initial distribution of
partons:
\begin{multline}
 T(Y) = \sum_{s=1}^\infty \frac{(-1)^{s-1} }{s!}  \sum_{n,\tilde n}
 \int d\B_{n} d\widetilde \B_{\tilde n} P(\B_n) \widetilde P(\widetilde \B_{\tilde n}) \\
 \int d\Z_s d\widetilde \Z_s f_s(y;\B_n|\Z_s)
 \tilde f_s(Y-y;\widetilde \B_{\tilde n}|\widetilde \Z_s ) \prod_i g({\bf z}_i - \tilde {\bf z}_i - \b)
\label{TST-full}
\end{multline}
Here $f_s(y;\B_{n}|\Z_{s})$ and $\tilde f_s(Y-y;\widetilde \B_{\tilde n} |\widetilde \Z_{s})$ stand
for the $s$-parton inclusive distributions obtained upon the evolution of the initial sets of
projectile-associated partons at positions $\B_n$ and of target-associated partons at
$\widetilde \B_{\tilde n}$. The straightforward way of calculations described above consists in
averaging over the sets $\B_{n}$ and $\widetilde \B_{\tilde n}$ prior to the integration over
$\Z_s$ and $\widetilde \Z_s$.

Another way of calculations can be realized making averaging over $\{n, B_{n}\}$ and
$\{\tilde{n},\widetilde B_{\tilde{n}}\}$ \emph{after} the convolution in $\Z_s$ and $\tilde Z_s$.
In other words, the Monte-Carlo averaging is done for the amplitude computed on the ``samples'',
particular realizations of the probability distributions which correspond to the set of inclusive
distributions $f_s$ and $\tilde f_s$. Each sample is generated by evolving a set of partons in
rapidity starting from the random number and initial positions of partons generated according to
the distributions  $P(n,\B_{n})$ and $\widetilde P(\tilde n,\widetilde \B_{\tilde n})$.

A set of $N$ partons with fixed coordinates $\widehat \X_{N}=\{ \hat {\bf x}_1,\ldots, \hat {\bf x}_{N}\}$ in
the transverse plane corresponds to the following set of the $s$-parton inclusive distribution
functions $f_s(Z_s)$ ($s=1\ldots N$):
\begin{equation}
 f_s(\Z_s) = \sum_{\{\hat {\bf x}_{i_1},..,\hat {\bf x}_{i_s}\}\in \hat \X_{N}}
\delta({\bf z}_1 - \hat {\bf x}_{i_1}) \ldots \delta({\bf z}_s - \hat {\bf x}_{i_s}). \label{fs-sample}
\end{equation}
The sum goes over all the \emph{ordered} subsets of $s$ variables out of $N$ which ensures the
normalization of $f_s$ and its symmetry properties. The same is for $\tilde f_s(\widetilde \Z_s)$
where the parton coordinates $\hat y_i \in \Y_{\tilde N}$. After  substituting (\ref{fs-sample})
the sample amplitude (cf. with \eqref{TST-full}) reads:

\begin{multline}
 T_{\rm sample} =  \sum_{s=1}^{N_{min}} \frac{(-1)^{s-1}}{s!}
 \sum_{\substack{ i_1,\ldots ,i_s =1 \\ i_m\neq i_l }}^{N} \
 \sum_{\substack{j_1,\ldots ,j_s =1 \\ j_m\neq j_l}}^{\tilde N}
 g(\hat {\bf x}_{i_1} - \hat {\bf y}_{j_1})\ldots g(\hat {\bf x}_{i_s} - \hat {\bf y}_{j_s}).
 \label{Tsample}
\end{multline}
Here $N$ and $\tilde N$ are the numbers of partons in the samples which come from the evolution of
the projectile- and the target-associated sets of partons and $N_{min}=\min(N, \tilde N)$ is the
minimal number of partons in these two samples.

The values
\begin{equation}
 g_{ij} \equiv g(\hat {\bf x}_i-\hat {\bf y}_j)
\end{equation}
computed on the samples of partons form a matrix which
can be looked upon as the adjacency matrix of the bipartite graph with weights $g_{ij}$.

For a given $s$ in the sum (\ref{Tsample}) there are $s!$ identical terms which correspond to permutations of the pairs $\{{\bf x}_i, {\bf y}_j\}$. This is since the summation in (\ref{Tsample}) is goes not only over the subsets $\{i_1,\ldots,i_s\} \in \{ 1, \ldots, N\}$ and  $\{j_1,\ldots,j_s\} \in \{ 1, \ldots, \tilde N\}$ but also over the permutations within these subsets due to symmetrized form of $f_s(\Z_s)$.
Taking this into account we arrive at the expression
\begin{equation}
 T_{\rm sample} =  \sum_{s=1}^{N_{min}} (-1)^{s-1}\sum_{i_1<i_2 \ldots < i_s} \sum_{j_1 , \ldots , j_s }
 g_{i_1 j_1}\ldots g_{i_s j_s}. \label{Tperm}
\end{equation}

It means that for each term of the sum with fixed $s$ one has to take all possible products of
$s$ elements of the matrix $G=\|g_{ik}\|$ so that the elements from the same line or column would not
enter same product. The products with the odd number of multipliers enter with $+1$, the ones with
the even -- with $-1$. Speaking in terms of mathematical definitions, this is a sum of the
permanents of all the square submatrices of the matrix  $G = \{g_{ij}\}$ with alternating sign. The
submatrices are obtained from the initial matrix $G$ by removing the relevant number of lines and
columns.

The permanent of a matrix (see e.g. \cite{Minc}) has a definition analogous to that of
the determinant, but without the alternation of the sign in the sum:
\begin{equation} \mbox{Per}\{A\} = \sum_{\pi\in S_n}\prod_{i=1}^{n} a_{i,\pi_i}  \label{Perm-def} \end{equation}
Despite its similarity with the determinant, the permanent is quite an inconvenient construction
for a numerical computation.
%For the size $n$ square matrix computing of its permanent  according to the definition (\ref{Perm-def}) takes about $n!$ operations.
The fastest known algorithm is given by the Ryser formula:

\begin{equation}
\mbox{Per}(A) = (-1)^n \sum_{S\subseteq\{1,\dots,n\}} (-1)^{|S|} \prod_{i=1}^n \sum_{j\in S} a_{ij},
\end{equation}
where summation goes over all the subsets $S$ of $\{1\ldots n\}$ and $|S|$ denotes the number of elements
in the subset. The number of operations for computing according to this formula is around
 $O(2^n n)$. This is much less than $O(n!)$ needed for the straightforward computation according to
(\ref{Perm-def}), however still incomparable with $O(n^2)$ operations needed for computing the
determinant.

One can easily generalize Ryser formula for the case of the amplitude $T_{\mathrm{sample}}$ in
(\ref{Tperm}):
\begin{equation}
 T_{\mathrm{sample}} = \sum_{s_1\subseteq\{1,\dots,N\}}
 \sum_{\substack{s_2\subseteq\{1,\dots,\tilde N\}, \\ |s_2|<|s_1|}} (-1)^{|s_2|-1}
 {\rm C}_{\tilde N -|s_2|}^{|s_1|-|s_2|} \prod_{i\in s_1}
\left( \sum_{j \in s_2} g_{ij}\right) ~.
 \label{TRyser}
\end{equation}
The estimated number of operations here is around $O(n 4^n)$ for a sample, which is much less than
the number of operations required for symmetrization of the full set of inclusive distributions
$f_s(\Z_s)$ on the grid in the straightforward way of the amplitude calculation.

Calculations based on the Monte-Carlo averaging of \eqref{TRyser} allow to to set an arbitrary
rapidity for the linkage point since the amplitude $T$ depends only on the overall rapidity, that
is the sum of the evolution times of two samples, as proven in \cite{Boreskov:2001nw}. The freedom
in the choice of the linkage point can be quite convenient for calculating the amplitude at large
rapidities when the Pomeron intercept  exceeds unity. In this case the number of partons in the set
grows exponentially with the evolution time and at some point the Monte-Carlo evolution of the set
can become very numerically expensive. So choosing the linkage point e.g. as a half of the overall
rapidity can in principle compensate the difficulties connected with the computing of the
permanents. Besides this, \eqref{TRyser} also serves as an expansion in the number of Pomeron
exchanges $|s_1|$ between the upper and lower blocks of the amplitude at the rapidity of the linkage
point.

If the application which we are using the stochastic approach for do not involve this expansion and
the overall rapidity is not too large, a much less numerically expensive method can be used. Taking
advantage of the independence of the amplitude \eqref{TST} on the linkage point we use the fact
that for zero evolution time the inclusive distributions $\tilde f_s(y=0,\Z_s)$ are proportional to
the particle--$m$Pomerons vertices (cf. \eqref{fs-vertex}) and hence are known. Substituting \eqref{fs-sample} into the
\eqref{TST} together with the known functions $\tilde f_s(y=0,\Z_s)$ and as before taking into
account the symmetry properties of $f_s(Y,\Z_s)$ in \eqref{fs-sample} we have:
\begin{equation}
T_{\rm sample} ({\bf b}) = \sum_{s=1}^{N} (-1)^{s-1} \tilde \mu_s \epsilon^s \sum_{i_1<i_2 \ldots <
i_s} \tilde p(\hat {\bf x}_{i_1} - {\bf b}, \ldots ,\hat {\bf x}_{i_s}-{\bf b}). \label{T-onesample}
\end{equation}
Here $\tilde \mu_s$ denotes the factorial moments of the distribution which corresponds to the set
$\tilde f_s(y=0,\Z_s)$ and $\tilde p_s$ stands for its coordinate dependent part normalized to
unity. In writing \eqref{T-onesample} we also assumed that the functions $g({\bf x})$ are narrow
and replaced them by $\epsilon \delta({\bf x})$. In case of no correlations in the vertex which for
example is the case for the eikonal vertices $\tilde p_s(\Z_s) = \prod p_{\tilde A}({\bf z}_i)$ and
\eqref{T-onesample} reads
\begin{equation}
T_{\rm sample} ({\bf b}) = \sum_{s=1}^{N} (-1)^{s-1} \tilde \mu_s \epsilon^s \sum_{i_1<i_2 \ldots <
i_s} p_{\tilde A}(\hat {\bf x}_{i_1} - {\bf b}) \ldots p_{\tilde A}(\hat {\bf x}_{i_s}-{\bf b}). \label{T-onesamplenocor}
\end{equation}
The amplitude is obtained as the Monte-Carlo average of \eqref{T-onesample} or
\eqref{T-onesamplenocor} over the sets of partons.

For the numerical calculations one must also fix the form of the distributions $p_{m_2}(b)$ and
$p_{\nu}(b)$ together with making a specific choice for the functions $g(b)$ in the convolution
(\ref{TST}). We take these distributions in the following form:
\begin{eqnarray}
p_{m_2} ({\bf b}) & = & m_2 \; \uptheta (a_{m_2} - |{\bf b}|);\\
p_{\nu} ({\bf b}) & = & \nu \; \uptheta (a_{\nu} - |{\bf b}|);\\
g ({\bf b}) & = & k \; \uptheta (a - |{\bf b}|); \label{g}
\end{eqnarray}
where we have introduced explicitly the interaction and convolution scales; the constants $\nu$ and
$m_2$ play a role of fusion and annihilation probabilities for a pair of partons per unit time
provided they are close enough in the transverse plane; $k$ is some appropriate numerical factor
\footnote{~ It allows to vary the scale $a$ preserving normalization of the function $g$,
 $\int d^2b g(\b) = k\, \pi a^2 = \epsilon$.}
 . The boost invariance condition (\ref{inv_cond}) in these notations reads:
\begin{equation}
 \lambda k \pi a^2 =  m_2 \pi a_{m_2}^2 + \tfrac{\nu}{2} \pi a_{\nu}^2
\end{equation}
while triple and quartic coupling constants are
\begin{equation}
\begin{array}{l} 
r_{3P} = \lambda \sqrt{\pi k} a = \sqrt{\tfrac{\pi}{k}} (m_2 a_{m_2}^2+ \tfrac{1}{2}\nu a_{\nu}^2)  / {a};\\
 \chi = \tfrac{\pi}{2} (m_2 a_{m_2}^2 + \nu a_{\nu}^2).
\end{array}
\end{equation}
For a given value of the triple coupling $r_{3P}$ by taking a higher value for the parameter $k$
one can release the scale $a$ and set it small enough compared to the characteristic scale of the
problem, which is, as mentioned, the condition of applicability of the stochastic approach. Though
three radii can in principle differ, in our numerical calculations for clarity we take all three
the same $a_{m_2} = a_\nu = a$. To reduce the number of free parameters in the model we also choose
$k=1$. The boost invariance condition for our choice takes the most simple form:
\begin{equation}
  \lambda =  m_2  + \nu / 2 ~. \label{inv-cond}
\end{equation}

The elastic amplitude is computed as a Monte-Carlo average of the expressions \eqref{Tsample} or \eqref{T-onesample} 
over the sets of partons which underwent stochastic evolution over the given time (rapidity) interval. The way of preparing these sets  is straightforward. First, the number of partons and their position in the transverse plane are generated acccording to the distribution defined by the particle--$m$-Pomeron vertices. After this, based on the position of partons in the transverse plane we define a probabilities $\gamma_i$ of splitting, death, fusion and annihillation of partons per unit time provided they preserve their positions. The survival probability (probability of having no change in the number of partons) as a function of time $y$ in this case is a simple exponential $p(y)=\exp(-\sum \gamma_i y)$. 

We generate time interval $\delta y$ before the change in the parton number according to the distribution $p(y)$. If the time interval does not exceed  $y_0=0.1 a/\sqrt{D}$ (10\% of the characteristic diffusion timescale) we define the type of the changing number event (toss it according to the total splitting, death, annihillation and fusion probabilities) and a candidate parton or a parton pair. Then we randomly change the positions of all the partons in the set according to the diffusion law with the time $\delta y$, do the necessary change in the number of partons and recalculate the number change probabilities $\gamma_i$ with the new parton number and the partons new coordinates.

If the time interval $\delta y$ exceeds $y_0$ we change the positions of partons in the set according to the diffusion law with the time $y_0$ and recalculate the probabilities $\gamma_i$. 

The steps are repeated until we reach the desired {overall} evolution time.

\section{The Pomeron propagator}
\label{sec4}

Now it is straightforward to attend to the numerical studies of the RFT. As a first example we
compare exact numerical results for the Pomeron propagator in zero-dimensional RFT and RFT with the
account of diffusion. The zero-dimensional case has been studied before both analytically in a series of works starting
from  \cite{Amati:1975fg} and numerically ( e.g. \cite{Braun:2006gy}) for zero value of the quartic
coupling. The most important result was the decrease of the two-point Green function or the Pomeron 
propagator (describing transition of one Pomeron into one Pomeron) with rapidity although 
the perturbative propagator has an exponential growth. It has also been 
reported in \cite{Bondarenko'06} that in the zero-dimensional theory with the $2\to 2$ coupling a constant asymptotic behaviour of the 
propagator is possible for some special choice of the quartic coupling, namely $\chi = r_{3P}^2/\Delta$ in our notations.

In terms of the stochastic model the propagator is the
single particle inclusive distribution $f_1(y, {\bf b})$ for the state which have evolved out of a
single parton at $y = 0$ (see \eqref{fs-vertex}).
The transition to zero-dimensional case is made in our numerical approach by setting the diffusion
coefficient equal to zero. Within our calculation scheme we investigate
the role of the quartic Pomeron coupling in the asymptotic behaviour of the propagator or
amplitude in both zero and two transverse dimensions Reggeon field theory.

We start with the zero dimensions case. Our aim is to illustrate the role of the enhanced diagrams
simply comparing the simple pole contribution to the propagator with the result of the full
numerical computation.

\begin{figure}[htbp]
\centering
 \includegraphics[width=0.8\hsize]{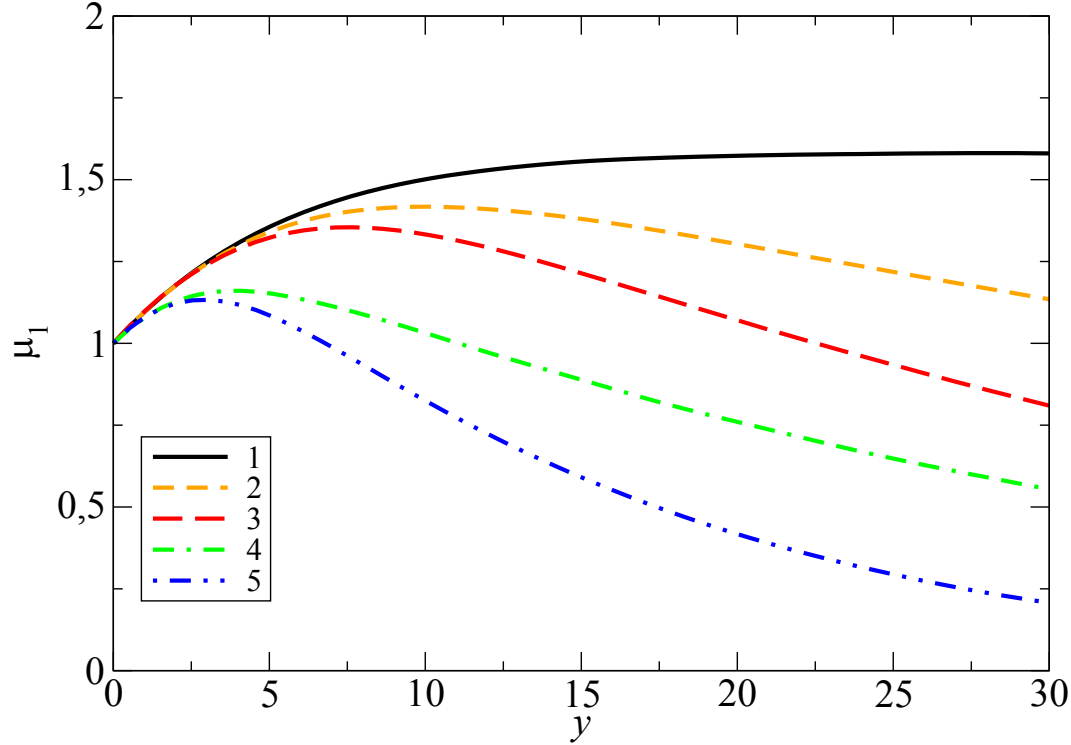}
\caption{The full Pomeron propagator in zero dimensions as a function of rapidity.  The stochastic
model parameters corresponding to the sets 1--5 are in the text.} \label{0D-plots}
\end{figure}

Here the propagator as a function of rapidity is simply the average number of partons as a function
of the evolution time. The evolution starts with a single parton. In fig.\ref{0D-plots} we present 
calculations for different sets of the parameters $\lambda$, $\nu$, $m_1$, $m_2$ (see
table\ref{table2}). The values of the parameters correspond to the same value of the intercept $\Delta
= \lambda - m_1 = 0.1$, satisfy the Lorenz invariance condition in zero dimensions $\lambda = m_2 +
\nu/2$ (cf. eq.(66) of \cite{Boreskov:2001nw}) and differ by values of the couplings $\r3P =
\lambda$ and $\chi = (m_2 + \nu)/2$ only.

\begin{table}[h]
 \caption{Parameters of the model and the RFT couplings}
 \vskip 2mm
\centerline{\begin{tabular}{|l|cccc|ccc|}
 \hline
Set  & $\lambda$ & $\nu$ & $m_1$ & $m_2$ & $\Delta$ & $\r3P$ & $\chi$ \\
 \hline
1 & 0.1 & 0.2 & 0 & 0 & 0.1 & 0.1 & 0.1\\
2 & 0.1 & 0.1 & 0 & 0.05 & 0.1 & 0.1 & 0.075 \\
3 & 0.1 & 0 & 0 & 0.1 & 0.1 & 0.1 & 0.05 \\
4 & 0.15 & 0.3 & 0.05 & 0 & 0.1 & 0.15 & 0.15 \\
5 & 0.15 & 0 & 0.05 & 0.15 & 0.1 & 0.15 & 0.075 \\
 \hline
 \end{tabular}}
 \label{table2}
\end{table}

As one can see from fig.~\ref{0D-plots}, the asymptotic behaviour of the propagator for
zero-dimensional theory strongly depends on the relative values of $m_2$ and $\nu$ or, in terms of
the RFT on the relation between the intercept and the couplings. The only possibility to have
constant asymptotic behaviour distinct from zero within the stochastic approach is to put $m_2=m_1=0$ as in the set~1. This
corresponds to equal triple and quartic coupling with $\Delta = \lambda$ (where $\Delta+1$ is the
Pomeron intercept) within  zero dimensional RFT which evidently satisfies the condition necessary for the constant propagator asymptotic behaviour 
outlined in \cite{Bondarenko'06}. The constant asymptotical value corresponds to the eq. (27) of 
\cite{Boreskov:2001nw} $\mu_1(\infty)=e/(e-1)\approx 1.582$. The same result is obtained by explicitly solving
numerically the differential equation (eq. (6) of \cite{Boreskov:2001nw}) for generating function
of the probability distribution.

Let us emphasize that all parameters of the stochastic model should be positive. Another strong
restriction is given by the Lorentz invariance condition which excludes some increasing regimes of
the stochastic model.

For the two dimensional case we take the same values of the stochastic model parameters as for the
zero-dimensional one. Additionally we introduce  the partonic interaction distance $a = 0.05$~fm
and the diffusion coefficient $\alpha' = 0.01$~fm$^{2} = 0.26$~GeV$^{-2}$. 
The Pomeron propagator now is a function
of both rapidity and impact parameter. The pole contribution or the bare propagator reads
\begin{equation}
 G(y,b) = \frac{1}{4 \pi \alpha' y} \exp\left(\Delta y -\frac{b^2}{4 \alpha' y}\right).
\end{equation}
It goes to delta function when the rapidity $y$ goes to zero and its value at $b=0$ is growing with
rapidity as $\sim e^{\Delta y}/y$.

\begin{figure}
 \centering
\includegraphics[width=0.8\hsize]{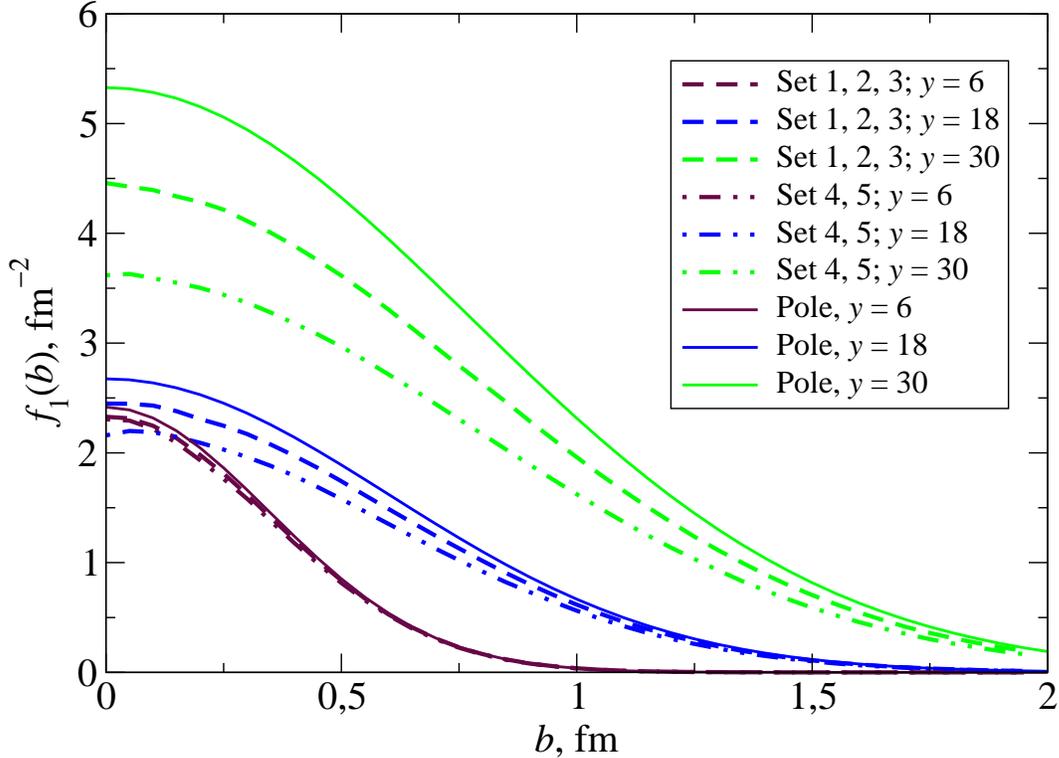}
\caption{The full Pomeron propagator as a function of impact parameter for different values of
rapidity $y$ in comparison with the perturbative one.}
\label{pprop-b}
\end{figure}

In figure~\ref{pprop-b} we show the calculation of the full propagator as a function of impact parameter
in comparison with the bare one for different rapidities.
The calculations show a negligible difference between sets of parameters which correspond to the
same value of the triple coupling $\r3P$ and  differ only by the value of the quartic coupling
$\chi$ ($\{1,2,3\}$ and $\{4,5\}$). This supports the statement of \cite{Abarbanel:1975me} that
presence of a finite number of higher than triple order terms in the RFT interaction Lagrangian may be ignored.

For making comparison with zero-dimensional case we take
the value of the propagator at $b=0$ multiplied by $\epsilon = \pi a^2$ 
which in terms of the stochastic model is the average number of partons with position within that area 
and on the other hand is the amplitude for a single-Pomeron exchange.
One may expect saturation in this quantity at sufficiently high
rapidities even in presence of diffusion, since, naively speaking, at low densities parton
splitting will prevail and at high densities fusion and annihilation will increase resulting in saturation eventually. However, as one
can see from fig.~3 the saturation is not reached within the range of rapidity achievable by the numerical 
calculation, unlike the case of zero transverse dimensions which is equivalent to zero slope. 
This is a consequence of the fact that the value of the slope which we used and which is close to the value dictated 
by the fits to the experimental data (e.g \cite{Ter-Martirosian}) is in fact quite large.

\begin{figure}
\centering
\includegraphics[width=0.8\hsize]{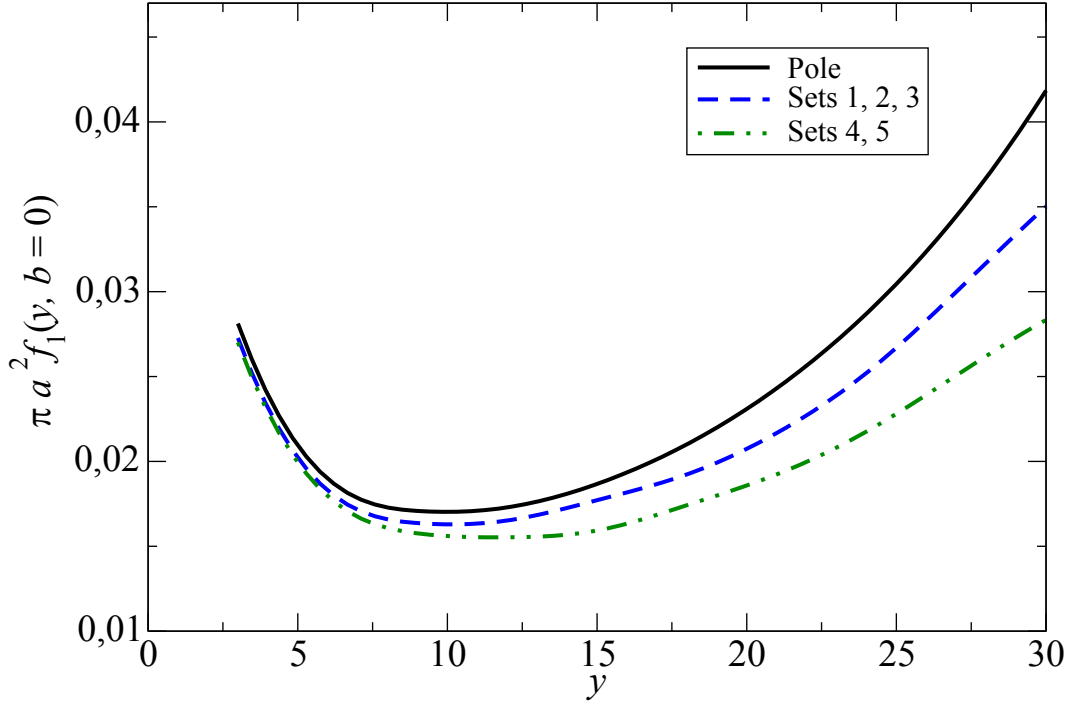}
 \caption{Full propagator scaled by $\epsilon=\pi a^2$ at $b=0$ as a function of rapidity in comparison with that of the bare
one. The sets of constants $\lambda$, $\nu$, $m_1$ and $m_2$ are the same as used for fig. 1.} \label{prop-zerob}
\end{figure}

The applicability of  zero dimensional approach depends on the interrelation of the parameters of
the Reggeon Field Theory. Within the stochastic model the applicability condition can be formulated
as follows. The number of splittings/fusions of partons must be already large for the rapidity $y_a$
which is required for the partons to diffuse to the distance of about the parton size.

That is, within the stochastic approach the regime similar to zero-dimensional case is realized for
\begin{equation}
 {\lambda a^2} \gg \alpha ' y_{a}.
\label{zero-d-cond}
\end{equation}

\begin{figure}
\centering
\includegraphics[width=0.85\hsize]{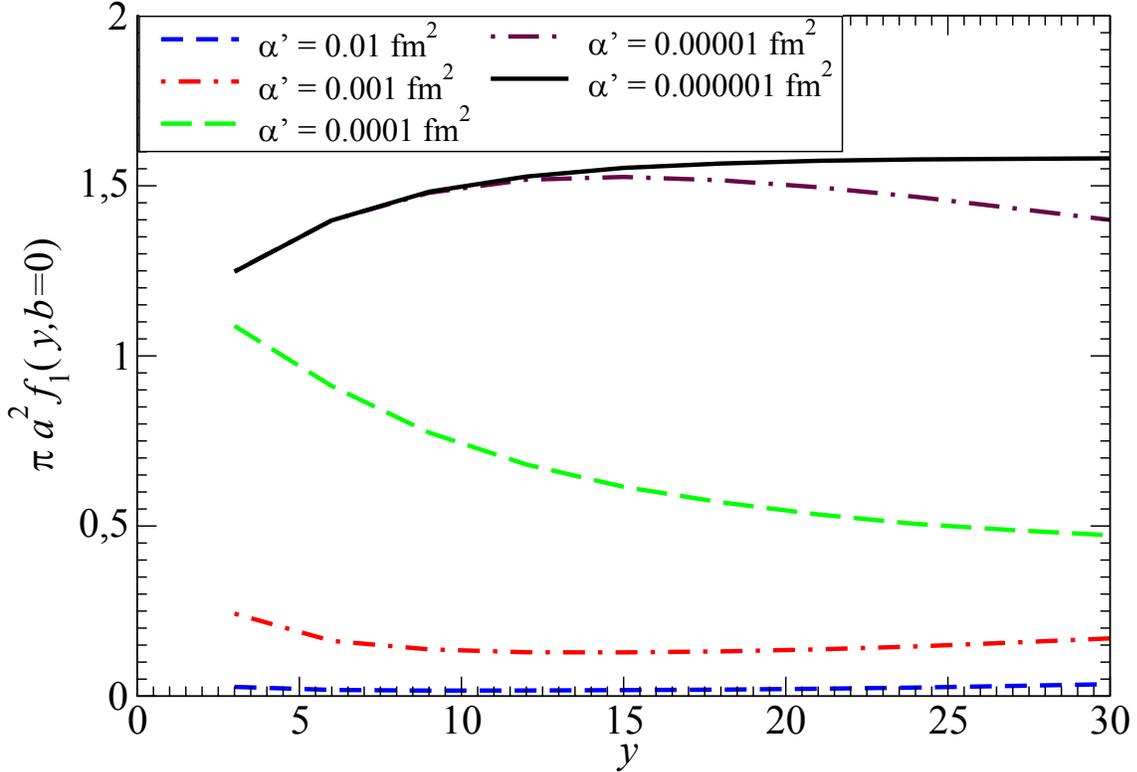}
 \caption{Full propagator scaled by $\epsilon=\pi a^2$ at $b=0$ for the constants $\lambda$, $\nu$, $m_1$ and $m_2$
of the set 1 and different slope $\alpha'$ values.} \label{prop-slope}
\end{figure}

This situation is illustrated in fig.~\ref{prop-slope} where we draw 
the same value as in fig.~\ref{prop-zerob} for the parameter set number 1. 
However this time the curves correspond to different values of the 
slope $\alpha'$ ranging from $0.01$~fm$^2$ to $10^{-6}$~fm$^2$ 
(from $0.26$ to $2.6\cdot 10^{-5}$~GeV$^{-2}$). One can easily see that the 
zero-dimensional regime is observed for $\alpha' = 10^{-5}$~fm$^2$ up to $y \simeq 10 $ 
which is in agreement with the estimate \eqref{zero-d-cond}. However, 
since the triple Pomeron coupling is comparatively small (and, consequently, 
$\lambda$ and $a$ are), it is rather the opposite limit that must be 
adopted for the applications and the diffusion must be fully taken into account.

%

%Another thing to observe in fig.~\ref{prop-slope} is the increased influence of the quartic coupling value on the propagator behavior 
%with decreasing of the slope value.

To conclude, in presence of diffusion the full Pomeron propagator at a fixed distance in the transverse
space {for the realistic slope value dictated by fits to the experimental data} is a growing function of rapidity within the reach of numerical calculations. {Within this rapidity range} it just
expresses a slower growth than the bare one (pole contribution). This is unlike the case with the
absence of diffusion where the full propagator drops at high energies or reaches constant value
comparatively fast. We also note that the behaviour of the propagator at large rapidities 
seem to depend on the slope value in a highly nontrivial way.  
{This points out an important role of the Pomeron slope for the set on of saturation and probable change to a subcritical behavior in the propagator.} Another important distinction between the zero-dimensional and 2-dimensional
theory is the role played by the quartic coupling. As one can see, the quartic coupling drastically
changes the propagator asymptotic behaviour in case of zero-dimensions, while in presence of diffusion its
role is negligible.

\section{Total and elastic cross sections.}
\label{sec5}

\subsection{Normalization of the amplitudes and the couplings.}

To proceed with applying the technique to the calculations of the cross sections we first fix the normalization of the
amplitude. This procedure is essential for the applications as it affects definitions of the
phenomenological couplings of the Pomeron to particle.

We use the elastic scattering amplitude $M(Y,\q)$ normalized in such a way that
\begin{equation}
 \sigma^{\rm tot} (Y) = 2\, \Im M(Y,\q=0) , \quad
 \sigma^{\rm el} =  \int \frac{d^2q}{(2\pi)^2}\, |M(Y,{\bf q})|^2 ~,
\end{equation}
where $Y=\ln (s/s_0)$, $s_0=1~\GeV^2$ and $\q^2=-t$.
The amplitude in the coordinate representation $f(Y,\b)$ is defined as
\begin{align}
 f(Y,\b) = \frac{1}{(2\pi)^2}\int d^2q\, e^{-i\q\b} M(Y,{\q}) ~.
\end{align}
With this definition of the Fourier transform, the perturbative Pomeron propagator in the coordinate representation coincides with the Green function of the diffusion equation with splitting. The cross sections are
\begin{equation}
 \sigma^{\rm tot} (Y) = 2\!\int d^2b\, \Im f(Y,{\bf b}) ~, \quad
 \sigma^{\rm el} = \int d^2b\, |f(Y,{\bf b})|^2 .
\end{equation}
We neglect the small real part of the Pomeron signature factor, so in what follows the Pomeron
amplitudes are considered as purely imaginary, $f(Y,\b) \simeq iT(Y,\b)$, $T\equiv\Im f$.

\begin{figure}[h]
 \centering
\includegraphics[width=0.6\hsize]{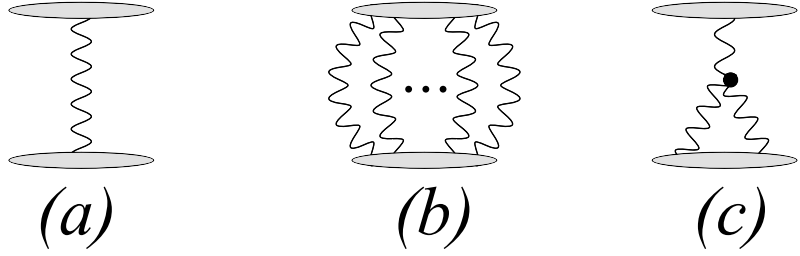}
\caption{Reggeon graphs used for normalization of the couplings.}
\label{rd}
\end{figure}

To establish the correspondence between the Pomeron--particle couplings and the distributions of projectile- and target-associated partons in the stochastic approach we use the contribution of single Pomeron exchange to the
total cross section (fig.\ref{rd}a). In the standard Regge parametrization the Regge-pole
contribution has the form
\begin{align}
 T_{\scs P}(Y,q^2) = \beta_{\Aph}(q^2)\beta_{\widetilde{A}}(q^2)\, e^{-\alpha'_P q^2 Y}e^{\Delta Y} ~,
\end{align}
where $\beta_{A/{\widetilde{A}}}$ are the couplings of the Pomeron to the colliding hadrons. They are often written in the form $\beta(q^2)=\beta_0 g(q)$ with $g(q=0)=1$. The most common choice for the profile $g$ is the Gaussian: $g(q) = \exp(-R^2q^2)$. In the impact
parameter representation one has
\begin{multline}
 T_{\scs P}(Y,b) = \beta_0^{\Aph} \beta_0^{\tilde A} \\ \int d^2 b_1
 d^2 b_2\, \widetilde{g}_{\Aph}\!(\b_1)\, G(Y,\b_2 - \b_1)\widetilde{g}_{\widetilde{A}}(\b-\b_2) ~,
\label{TP_b}
\end{multline}
where $\widetilde{g}_{A/\widetilde{A}}$ are the Fourier images of the couplings profiles normalized to unity
\begin{align}
 \widetilde{g}_{A/\widetilde{A}}(\b) = \frac{1}{(2\pi)^2}\int d^2 q e^{-i\q\b} g_{A/\widetilde{A}}(\q) ~,
\end{align}
and
\begin{align}
 G(Y,b) = \frac{\exp(-\tfrac{b^2}{4 \alpha'Y})}{4\pi\alpha' Y} e^{\Delta Y}
 \label{GP}
\end{align}
is the Pomeron propagator $\exp((\Delta-\alpha' \q^2)Y)$ in the coordinate representation.

To model a single Pomeron exchange in the stochastic model approach we set $m_2=\nu=0$ and take
only the term with $s=1$ from the sum (\ref{Tperm}) (this corresponds to $|s_1|=1$ terms in
(\ref{TRyser})). For the imaginary part of the amplitude using (\ref{TST}) and (\ref{g}) we
evidently get
\begin{multline}
 T_{\scs P}(Y,b) = 
\epsilon N_{\Aph} N_{\widetilde{A}}\,\\
 \int d^2 b_1 d^2 b_2\, p_{\Aph}\!(\b_1)\, G(Y,\b_2 - \b_1) p_{\widetilde{A}}(\b-\b_2) ~,
 \label{TP}
\end{multline}
where $N_{A/{\widetilde{A}}}$ stand for the average number of partons attributed to the projectile
$A$ and target $\widetilde{A}$ at the start of evolution and $p_{A/{\widetilde{A}}}(b)$ for their
distributions in the transverse plane which are normalized to unity.
 The function $G(Y;\b)$
describes diffusion of partons in the transverse plane and growth of their number with rapidity. It
coincides exactly with the Pomeron pole propagator \eqref{GP} if the diffusion coefficient
$D=\alpha'$. It follows from comparison of eqs.\eqref{TP_b} and \eqref{TP} that the initial parton
distributions in the transverse plane coincide with the Pomeron-particle coupling profile in the coordinate representation and the number of partons is proportional to the value of the coupling at zero momentum transfer.
\begin{align}
 \widetilde{\beta}^{A/\widetilde{A}}_0 \tilde g_{A/\widetilde{A}} (b) = \sqrt{\epsilon} N_{A/{\widetilde{A}}}\, p_{A/{\widetilde{A}}}(b) ~.
\end{align}

The total cross section for the proton--proton case where $N_{\Aph}=N_{\widetilde{A}}\equiv N$ is
correspondingly
\begin{equation}
 \sigma_{\scs P}^{\rm tot} =2 \beta^2 e^{\Delta y}. \label{def-tot}
\end{equation}
where the constant $\beta \equiv \widetilde{\beta}_0^p= N \sqrt{\epsilon}$.

Now it is straightforward to write projectile- and target-associated parton distributions in the
quasieikonal approximation. The quasieikonal amplitude is written as an expansion in the number $s$
of Pomeron exchanges with a special choice for the particle--$s$-Pomeron vertices. In the
normalization of the amplitude which we have adopted it reads:
\begin{equation}
 T(Y,{\bf b}) = \sum_{s=1}^\infty T_{{\scs P}}^{(s)}(Y,{\bf b}) = \sum_{s=1}^\infty \frac{(-1)^{s-1} C^{s-1}}{s!}
 \left(T_{{\scs P}}(Y,{\bf b})\right)^s. \label{T-qeik}
\end{equation} The eikonal approximation corresponds to the special choice of $C=1$.
Writing down the $s$ Pomeron exchange contribution for $Y\to 0$ explicitly and taking into account
\eqref{TST}, one finds that it corresponds to the convolution of two $s$-parton inclusive
distribution of the following form:
\begin{equation}
 \epsilon^{s/2} f^{A/\tilde A}_s({\bf b}_1,\ldots, {\bf b}_s) = \left(\beta_0^{A/\tilde A}\right)^s C^{(s-1)/2}\prod_{i=1}^s p_{A/ \tilde A}({\bf b}_i).
\end{equation}
This corresponds to the partons independently distributed in the transverse plane with the
distribution $p_{A, \tilde A}({\bf b}_i)$ and parton number distribution with the factorial moments
$\mu_s = \left(\beta_0^{A/\tilde A}\right)^s C^{(s-1)/2}/\epsilon^{s/2}$. This set of the factorial
moments $\mu_s$ in its turn corresponds to the ``quasipoissonian'' distribution (poissonian
distribution for the eikonal) with
\begin{eqnarray}
& P_N = \kappa^{N-1} \frac{\gamma^N}{N!}e^{-\gamma \kappa}, \quad n=1,\ldots,\infty &\\
& P_0 =  1 - \sum_{N=1}^\infty P_N = 1-\frac{1}{\kappa}(1-e^{-\gamma \kappa}) ~,&
\end{eqnarray}
where $\kappa = \sqrt{C}$ and $\gamma = \cfrac{\beta_0^{A/\tilde A}}{\sqrt \epsilon} $.

It is interesting to note that the quasieikonal approximation \eqref{T-qeik} can be reproduced
within the stochastic approach for any rapidity. This is provided if the relation between the inclusive
distributions, $f_s(y;{\bf b_1}, \ldots, {\bf b}_s) = C^{(s-1)/2} \prod f_1(y;{\bf b}_i)$, which valid at $y=0$ due to the 
independent distribution of partons in the transverse plane is also
fulfilled for any rapidity value.
This amounts to the factorial moments of the parton number distribution growing as
\begin{equation}
 \mu_s(y) = e^{\Delta y}\mu_s(y=0)
\end{equation}
and (for the Gaussian initial profile) independent Gaussian distribution of parton coordinates in
the transverse plane with standard deviation growing along with rapidity as
\begin{equation}
 R^2(y) = R^2 + \alpha' y.
\end{equation}

To fulfill these conditions in the course of computation it is necessary not only to forbid fusion
and annihilation of partons but to impose an additional condition on the level of linking of the
inclusive distributions $f_s(\Z_s)$. Namely, all partons which coordinates enter the set $\Z_s$
must belong to different connected components, that is, must be produced from different initial
partons.  In our numerical scheme this amounts to taking only such contributions to the sum in the
(\ref{fs-sample}) which do not contain coordinates of the partons from the same connected
component.

The corresponding numerical procedure is realized by dividing the matrix $G = \|g_{ik}\|$ entering
(\ref{Tsample}) into blocks which correspond to interaction of partons belonging to the same
connected components. The blocks are numbered by connected components of the graph which come from
$y=0$ and $y=Y$. Terms entering the sum in (\ref{Tsample}) are those whose multipliers come from
different blocks with different components number, each number of component can not be used twice
in same product. In other words, all the multipliers must come from different block lines and
different block columns. Effectively this corresponds to substituting the matrix $G$ by $\tilde G$
which dimensions are the numbers of connected components and elements being sum of the elements of
the matrix $G$ which enter the corresponding block. The matrix $\tilde G$ is processed according to
the usual procedure described in the section 3.

\begin{figure}
\centering
\includegraphics[width=0.5\hsize]{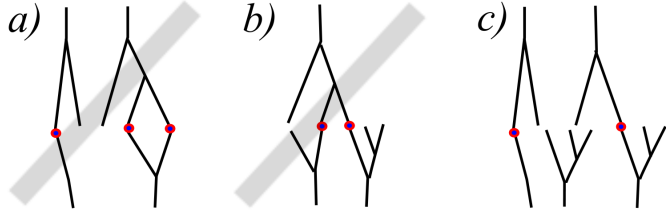}
\caption{Schematic view of the contribution entering the quasieikonal amplitude in the stochastic
approach. Only graphs of type $c)$ contribute.} \label{graph-qeik}
\end{figure}

The contributions which enter the amplitude (\ref{Tsample}) computed on the partonic samples can be
schematically illustrated by the fig.~\ref{graph-qeik} where solid lines correspond to parton
propagation in time. Obviously, in the quasieikonal approximation we sum up only a certain part of
the contributions to the amplitude. However the result is obviously independent from the position of the partonic samples
linkage point in rapidity, the same as for the full calculation with
account of fusion and annihilation.

As the next step we establish a link of the triple Reggeon coupling, $\r3P = \lambda \sqrt{\epsilon}$ in our
notations, with the relevant quantities used in fits to the experimental data by different groups
(see \cite{Kaidalov:1979jz}, \cite{Goulianos'95+'99}, \cite{Martin+Luna'09}). To do this we write
explicitly the expression for the diffractive cross section in the lowest order in parton
splitting, which is expressed via the single triple interaction contribution to the amplitude (see
fig.\ref{rd}c). Within the stochastic model we start with the contribution to the amplitude
in the Schwimmer approximation with $\alpha' \ne 0$ which is first order in $r_{3P}$.
The Schwimmer approximation implies eikonal vertices for the target and $m_2=\nu=0$. In this case it is straightforward to write down a solution to the set of equations (43) of \cite{Boreskov:2001nw} for the inclusive distribution. The convolution of $f_2(Y, b_1, b_2)$ with the particle -- two Pomeron vertex gives the stochastic model expression for the amplitude in the impact parameter representation:
\begin{multline}
 T_{3P}(Y,b) \equiv \int_{y_{\rm min}}^Y  dy \tilde T_{3P}(y,Y;b)=\\
 - \lambda \epsilon^2 N_{\Aph} N_{\tilde{A}}^2
 \int_{y_{\rm min}}^Y dy\int d^2b_1 d^2b_0 d^2b_2 d^2b_2' \,
 p_A(\b_1) G(y;\b_0-\b_1) \\
 \times G(Y-y;\b_2-\b_0) G(Y-y;\b_2'-b_0) p_{\tilde{A}}(\b-b_2) p_{\tilde{A}}(\b-b_2') ~.
\end{multline}
Its discontinuity which corresponds to the intermediate state with rapidity gap $y$ (diffractive
dissociation) is of the positive sign and double magnitude.
The corresponding single diffractive cross section is
\begin{align}
 \frac{d\sigma^{\rm SD}}{dy_M} &\equiv M^2\frac{d\sigma^{\rm SD}}{dM^2}
 = - 2 \int d^2b \, \tilde T_{3P}(y_M,Y;b) = \nonumber \\
 &= \r3P \beta_{\Aph}(0) e^{\Delta (Y-y)} \int \frac{dt}{2 \pi}\,\left(\beta_{\tilde{A}}(t)\right)^2
 e^{2(\Delta+\alpha't)y} ~.
 \label{def-SD}
\end{align}
where $y_M=Y-y=\ln(M^2/s_0)$.

We make direct comparison of (\ref{def-tot}) and (\ref{def-SD}) with the corresponding definitions
of \cite{Kaidalov:1979jz} ($r_K$), \cite{Goulianos'95+'99} ($r_G$) and \cite{Martin+Luna'09} ($r_L$) to
find out that
 $$\r3P = \sqrt{\pi}\, r_K = \sqrt{\frac{1}{2\pi}}\, r_L = \frac{r_G}{2\sqrt{2}} ~.$$

\subsection{The cross sections.}

In order to estimate the role of loop corrections it is instructive to compare our calculations with
one of the realistic fit to the experimental data. Our aim is not to fit the data better -- we want to
estimate the relative contribution of the loop diagrams at realistic parameter values.
 As a starting point for the numerical calculation of the cross section we take a quasieikonal fit
to data according to \cite{Ter-Martirosian} with
\begin{equation}
 T(Y,{\bf b}) = \sum_{n=1}^{\infty} \frac{(-C)^{n-1}}{n!} \left( T_P(Y,{\bf b})\right)^n ~,
 %\frac{\left[-\omega(Y,{\bf b})\right]^n}{n!}
\end{equation}
where
\begin{equation}
 T_P(Y,\b) = \frac{g_0^2 \exp(\Delta Y)}{R_P^2 + \alpha'Y} \exp[-\tfrac{1}{4} b^2/(R_P^2+\alpha'Y)].
\end{equation}
The fit has a Gaussian parametrization of the Pomeron--particle vertex. The parameters of the fit
are as follows:
\begin{align}
\nonumber  \Delta &= 0.12 ~, \\
\nonumber g_0^2 & =  2.14~\text{GeV}^{-2} \approx 0.083 \fm^2 ~, \\
\alpha'_P & = 0.22~\GeV^{-2} \approx 0.0085 \fm^2 ~,  \label{qeik-param}\\
\nonumber R_P^2 & = 3.30 \GeV^{-2} \approx 0.128 \fm^2 ~,\\
\nonumber C & = 1.5 ~.
\end{align}
We take initial parton distribution in the transverse plane according to this fit in the Gaussian form accounting for a
different amplitude normalization in \cite{Ter-Martirosian}
 \footnote{In paper \cite{Ter-Martirosian} one uses normalization $\sigma^{\rm tot}(s)= 8\pi \Im T(s,\q=0)$.}:
\begin{align*}
  p_p(\b) = \frac{1}{4\pi R^2} \exp\left[-\frac{b^2}{2 R^2}\right]
\end{align*}
and the factorial moments of the parton number distribution as
\begin{align*}
\mu_s = \left( \bar N \right)^s C^{\frac{s-1}{2}}; \quad \bar N =\frac{2g_0}{a}
\end{align*}
We fix the Pomeron intercept $\Delta=\lambda - m_1=0.12$ and the triple Pomeron coupling
 $\r3P = \lambda \sqrt{\pi a^2} = 0.087$~GeV$^{-1}$ which corresponds to the value $r_K = 0.05\GeV^{-1}$
from \cite{Kaidalov:1979jz} in our normalization.

Our aim is to illustrate dependencies of the cross sections on the parton size (linkage distance)
$a$ and quartic coupling $\chi$. Following the relations outlined in table \ref{RFT-ST-rel} and
throughout the section we define several sets of the partonic scheme parameters which correspond to
the parameter set \eqref{qeik-param} and differ with transverse size of the parton $a$ and the
values of the quartic coupling $\chi$. We fix the parton size values at the level of 5\% and 10\%
of the proton radius. The quartic couplings depending on relation between $\lambda$ and $m_2$ for
the sets are $\chi_1=\chi_4 = 0.0005569$~fm$^2$, $\chi_2 = 0.0002785$~fm$^2$, $\chi_3 =
0.0011134$~fm$^2$. The sets of parameters are presented in table~\ref{scheme_param}.

\begin{table}[h]

 \caption{Parameters used for the full calculation within the stochastic scheme. 
 For all the sets $R = 0.36$~fm and $D=0.008538$~fm$^2$ are taken.}
 \vskip 2mm
\centerline{\begin{tabular}[]{|c|c|c|c|c|c|c|}
 \hline
 Set & $a$,~fm & $\lambda$ & $m_1$ & $m_2$ & $\nu$ & $\bar N$ 
\\
\hline
  1 & 0.018 & 0.54722 & 0.42722 & 0 & 1.09488  & 32.02  
\\
  2 & 0.018 & 0.54722 & 0.42722 & 0.54722 & 0   & 32.02   
\\
  3 & 0.036 & 0.27361 & 0.15361 & 0 & 0.54722  & 16.01 
\\
  4 & 0.036 & 0.27361 & 0.15361 & 0.27361 & 0  & 16.01  
\\
 \hline
\end{tabular}}
\label{scheme_param}
\end{table}

\begin{figure}[!t]
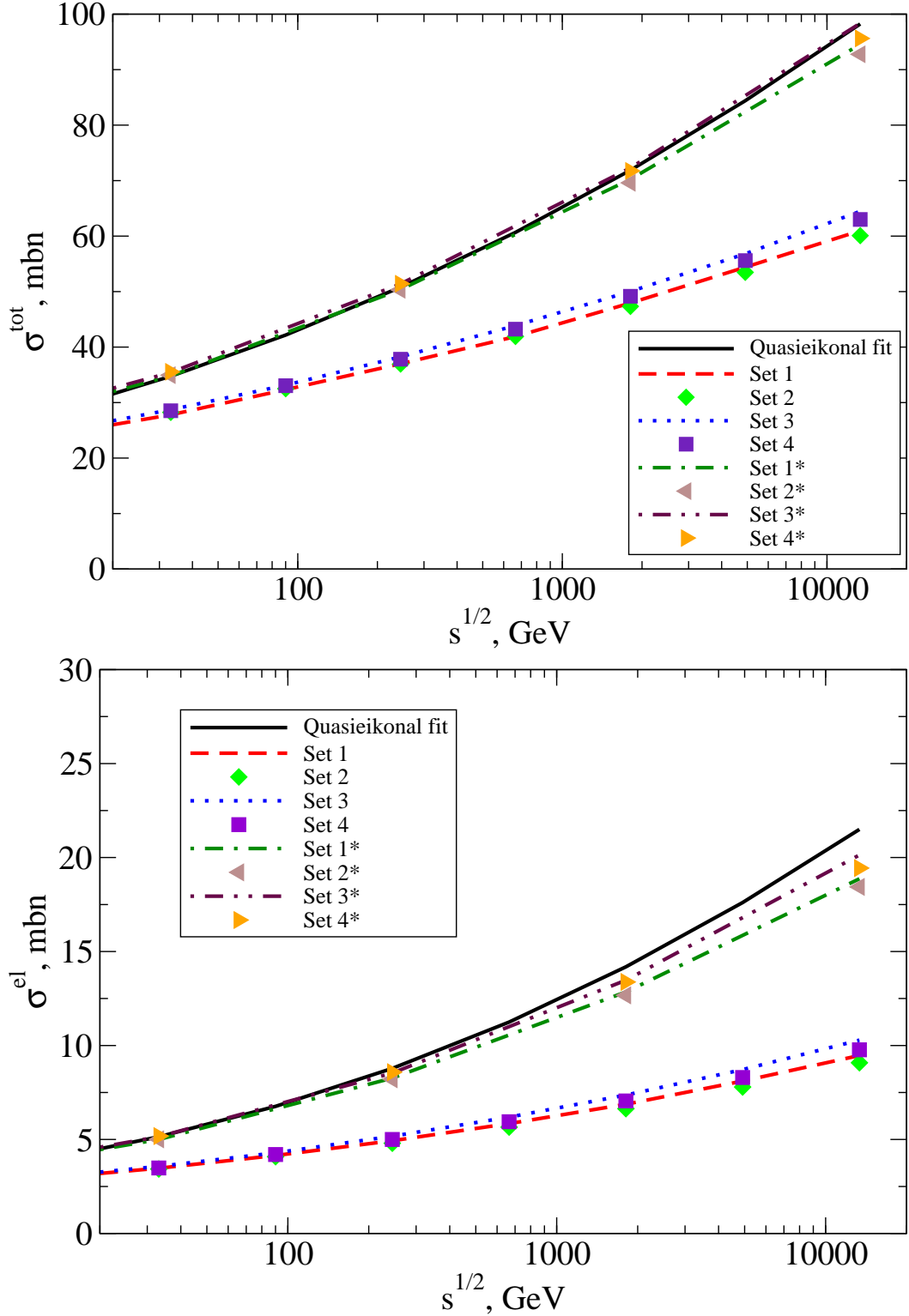

 \centering
\includegraphics[width = 0.8\hsize]{fig7a-stot.eps} \includegraphics[width = 0.8 \hsize]{fig7b-sel.eps}
\caption{Total and elastic cross sections; full calculation compared to the quasieikonal fit \cite{Ter-Martirosian}.}
\label{fig-014}
\end{figure}

The results of the full calculation are presented in figure~\ref{fig-014}. As one can see, Pomeron
interaction leads to a certain shadowing in the amplitudes and hence slows down the cross section
growth compared to the quasieikonal approximation. The value of the cross section is mainly defined
by the regularization scale (parton size $a$) and not by the value of the quartic coupling, the
same as for the Pomeron propagator. We have also checked numerically the independence of the result
on the linkage point and found it to hold within 2\% accuracy. We have also verified that 
the quasieikonal approximation is reproduced numerically in our computation once we impose restrictions 
on the parton sets evolution and their linkage procedure as described in section 5.1.

It is possible by changing parameters to bring the results of the full calculation close to the
quasieikonal fit to data. We did so by leaving all the coupling the same and increasing the
intercept value simply reducing $m_1$.  As one can see in the fig. \ref{fig-014} changing the value
of $\Delta=\lambda -m_1$ to 0.165 brings the full calculation results close to the original
quasieikonal fit ($\Delta=0.12$). The curves for the derived sets with $\Delta=0.165$ are
additionally marked with an asterix (1*, 2*, 3*, 4*) with respect to the original sets with
$\Delta=0.12$. The change in the value of the quartic coupling $\chi$ has a minor influence on the
cross section, much less than a change in the regularization scale (parton size) $a$.

The same as for the intercept value 1.12, the quartic coupling value has a minor influence on the values of the cross sections. For the energies reachable in the nearest future changing quartic coupling by a factor of 2 amounts to less than 5\% change in the cross sections.

\section{Conclusion.}

We  developed a numerical realization of the RFT
which allows to obtain the exact value of scattering amplitude taking all loops into account
using the equivalence between the Reggeon Field Theory with Pomeron scattering term and
parton stochastic model (reaction-diffusion approach).
Unlike other calculations with account of loops for no-zero Pomeron slope \cite{Martin+Luna'09,Kaidalov:2009aw,Ostapchenko:2010gt} our aproach does not involve infinite number of fine-tuned couplings for transitions of $m$ into $n$ Pomerons where results depend strongly on the chosen $n,m$ dependencies of the vertices.
Our calculations includes triple and quartic couplings only and, to our knowledge, is the first numerical all-loop calculation for the case of non zero slope.
This calculation possesses the expected properties of the approach: boost invariance and
projectile-target symmetry. 
Moreover, it gives the exact amplitude as an expansion in the number of the Pomeron exchanged at a given rapidity value $y$
which provides an straightforward opportunity of calculating the quantities which involve this expansion, 
such as multiplicity distribution in the  narrow rapidity interval.

The stochastic model has several parameters, most of them have direct correspondence within RFT and
are fixed through its couplings. One of the two parameters which are not fixed at once is the
parton interaction distance $a$. It is equivalent to the ultraviolet cutoff or the Pomeron size for
the calculations within the RFT thus playing a role of the RFT renormalization scale. Another
parameter is the ratio of parton fusion to parton annihilation probabilities. This ratio defines the
relative value of triple and quartic Pomeron couplings and up to our knowledge there are no experimental
constraints on its value yet.
However, as we show numerically the role of the quartic coupling in the RFT with the slope $\alpha'$
distinct from zero is negligible for the asymptotic behaviour of the amplitudes and the cross sections
within the reach of numerical computations. This is unlike the case of the zero slope.

As an illustration of use of our approach we compute the total and elastic proton-proton cross sections
in the approximation of quasieikonal proton-Pomeron vertices. Comparison of the full calculation with the quasieikonal fit
shows that for the relevant energy range  the full account of the Pomeron interactions can be effectively imitated in the quasieikonal 
approximation by reducing the intercept from $\Delta=0.165$ to $\Delta=0.12$. 

We hope that our study will be helpful in understanding the role of relative value of the Pomeron
loop contribution in high energy hadronic scattering.

\section*{Acknowledgement.}
Authors want to thank O.~Kancheli and S.~Ostapchenko for discussions and helpful remarks. 
R.K. and L.B. acknowledge support by the NFR, Project 185664/V30.  
Work of R.K. was also supported by the RFBR grant 09-02-01327-a.
K.B. acknowledges financial support from CRDF, grant RUP2-2961-MO-09.
R.K. is thankful to the family of his aunt Larisa Kirillova for the hospitality in Moscow.

\end{document}